\documentclass[a4paper,11pt]{article}
\usepackage{pos}

\title{Quantum Simulation of Finite Temperature Schwinger Model via Quantum Imaginary Time Evolution}
\ShortTitle{Quantum Simulation of Finite-T Schwinger Model via QITE}

\author*[a]{Juan W. Pedersen}
\author[b,c]{Etsuko Itou}
\author[c,d,e]{Rong-Yang Sun}
\author[d,e,f,g]{Seiji Yunoki}

\affiliation[a]{Graduate School of Arts and Sciences, University of Tokyo,
Komaba, Meguro-ku, Tokyo 153-8902, Japan}

\affiliation[b]{Yukawa Institute for Theoretical Physics, Kyoto University, Kitashirakawa Oiwakecho, Sakyo-ku, Kyoto 606-8502, Japan}

\affiliation[c]{Interdisciplinary Theoretical and Mathematical Sciences Program (iTHEMS), RIKEN, Wako, Saitama 351-
0198, Japan}

\affiliation[d]{Computational Materials Science Research Team, RIKEN Center for Computational Science (R-CCS), Kobe, Hyogo 650-0047, Japan}

\affiliation[e]{Quantum Computational Science Research Team, RIKEN Center for Quantum Computing (RQC), Wako, Saitama 351-0198, Japan}

\affiliation[f]{Computational Quantum Matter Research Team, RIKEN Center for Emergent Matter Science (CEMS), Wako, Saitama 351-0198, Japan}

\affiliation[g]{Computational Condensed Matter Physics Laboratory, RIKEN Cluster for Pioneering Research (CPR), Saitama 351-0198, Japan}

\emailAdd{pedersen@hep1.c.u-tokyo.ac.jp}
\emailAdd{itou@yukawa.kyoto-u.ac.jp}
\emailAdd{sun-rongyang@outlook.com}
\emailAdd{yunoki@riken.jp}

\abstract{
We study the Schwinger model at finite-temperature regime using a quantum-classical hybrid algorithm.
The preparation of thermal state on quantum circuit presents significant challenges.
To address this, we adopt the Thermal Pure Quantum (TPQ) state approach and apply the Quantum Imaginary Time Evolution (QITE) algorithm to implement the necessary imaginary time evolution.
We first compute the chiral condensate in the massless Schwinger model, verifying its consistency with the analytical solution.
We then simulate the massive Schwinger model with non-zero topological $\theta$-term to investigate the temperature and $\theta$-dependence of the chiral condensate.
Our method works well even at non-zero $\theta$ regime, while the  conventional lattice Monte Carlo method suffers from the sign problem in this system.
}

\FullConference{The 40th International Symposium on Lattice Field Theory (Lattice 2023)\\
July 31st - August 4th, 2023\\
Fermi National Accelerator Laboratory\\}

\begin{document}
\maketitle

%%%%%%%%%%%%%%%%%%%%%%%%%%%%%%%%%%%%%%%%%%%%%%%%%%%%%%%%%%%%%%%%%%%%%%%%%%%
%%%%%%%%%%%%%%%%%%%%%%%%%%%%%%%%%%%%%%%%%%%%%%%%%%%%%%%%%%%%%%%%%%%%%%%%%%%
\section{Introduction}

Quantum simulation of quantum field theories at finite-temperature regime is a challenging endeavor. 
This complexity is mainly due to the nature of thermal state as a mixed state, which complicates its preparation on quantum circuit.
A possible approach to this problem is usage of a ``typical pure state'', the so-called Thermal Pure Quantum (TPQ) state~\cite{Sugiura_2012,Sugiura_2013}.
In thermal equilibrium system, the TPQ state is defined as one whose expectation value approximates the thermal ensemble average of local observables at finite volume.
It is proven that the expectation value calculated with the TPQ state converges to the thermal ensemble average in the thermodynamic limit~\cite{Sugiura_2012}.
Although this theoretical framework was initially introduced as a novel formulation of statistical mechanics, it also has the potential for use in the quantum simulation of gauge theories. For example, its applicability to $Z_2$ gauge theory has been discussed in Ref.~\cite{davoudi2022quantum}.

We use a specific realization of the TPQ state termed as the canonical TPQ state~\cite{Sugiura_2013}. This state is defined by applying the imaginary time evolution operator to random state. Implementing imaginary time evolution on quantum circuit is also challenging since it is expressed by non-unitarity operators. 
Several algorithms have been proposed, mainly to address the ground state energy problem, such as variational methods~\cite{McArdle_2019}, probabilistic methods~\cite{Liu_2021}, power methods~\cite{seki2021}, and the Quantum Imaginary Time Evolution (QITE) algorithm~\cite{Motta_2019}.
Here, we employ the QITE algorithm combining the random-state generation algorithm to generate the TPQ state.

In this work, we simulate the temperature-dependence of the chiral condensate of the Schwinger model using the TPQ state and the QITE algorithm.  We perform the massless case to confirm the validity of our method by comparing our results with the analytic result. Furthermore, we explore the massive and non-zero $\theta$ regime where the sign problem appears in the conventional Monte Carlo method.
%%%%%%%%%%%%%%%%%%%%%%%%%%%%%%%%%%%%%%%%%%%%%%%%%%%%%%%%%%%%%%%%%%%%%%%%%%%
%%%%%%%%%%%%%%%%%%%%%%%%%%%%%%%%%%%%%%%%%%%%%%%%%%%%%%%%%%%%%%%%%%%%%%%%%%%
\section{The Schwinger model}
Let us first briefly overview the Schwinger model.
The Schwinger model is a 1+1 dimensional U(1) gauge theory. The qubit description of the Schwinger model Hamiltonian~\cite{Honda:2021aum} is given by 
\begin{equation}
  H=\frac{1}{4 a} \sum_{n=1}^{N-1}\left[X_{n} X_{n+1}+Y_{n} Y_{n+1}\right]+\frac{m}{2} \sum_{n=1}^{N}(-1)^{n} Z_{n}+\frac{a g^{2}}{2} \sum_{n=1}^{N-1}\left[\sum_{i=1}^{n-1} \frac{Z_{i}+(-1)^{i}}{2}+\frac{\theta}{2\pi}\right]^{2},\label{eq:hamiltonian_spin}
\end{equation}
where $N,a$ and $g$ denote the number of lattice site, the lattice spacing, and the coupling constant, respectively. $X_i, Y_i, Z_i$ are the Pauli matrices.
Here, we solve the Gauss's law constraint and impose the open boundary condition to remove the degree of freedom for the U($1$) gauge field.
Note that the non-local terms are induced due to the Gauss's law constraint in the last term in Eq.~\eqref{eq:hamiltonian_spin}. The topological $\theta$-term is shown as a constant shift of the electric-field ($Z_i$), and thus no additional difficulty emerges by the insertion in this formula. 

In this work, we investigate the (discrete) chiral symmetry as a function of temperature. The order parameter is given by the chiral condensate, $\langle\bar{\psi}\psi\rangle$, where $\psi$ is a fermionic field operator.
The quantity can be also expressed by the Pauli matrices using the Jordan-Wigner transformation as follows:
\begin{equation}
  \langle\bar{\psi} \psi\rangle=\frac{1}{N} \sum_{i=1}^{N} (-1)^{i}\frac{1+\left\langle Z_{i}\right\rangle}{2a}.
  \label{eq:chiral_condensate}
\end{equation}
In the massless case, its temperature-dependence has been derived analytically based on the Lagrangian formalism
~\cite{sachs2010finite} as
\begin{equation}
  \langle\bar{\psi} \psi\rangle=-\frac{m_{\gamma}}{2 \pi} e^{\gamma} e^{2 I\left(\beta m_{\gamma}\right)} \quad,\quad I(x)=\int_{0}^{\infty} \frac{1}{1-e^{x \cosh (t)}} d t ,\label{eq:exact_SW}
\end{equation}
where $\gamma$ is the Euler constant and $m_{\gamma}$ is the mass gap $m_\gamma =g / \sqrt{\pi}$.
Note that at zero temperature, it takes the value 
$\langle\bar{\psi} \psi\rangle \simeq-0.1599$~\cite{Gross_1996}.
On the other hand, it is difficult to solve this model analytically in massive regime.
Furthermore, when dealing with non-zero $\theta$ regime, the conventional lattice Monte Carlo method is not doable due to the sign problem.

%%%%%%%%%%%%%%%%%%%%%%%%%%%%%%%%%%%%%%%%%%%%%%%%%%%%%%%%%%%%%%%%%%%%%%%%%%%
%%%%%%%%%%%%%%%%%%%%%%%%%%%%%%%%%%%%%%%%%%%%%%%%%%%%%%%%%%%%%%%%%%%%%%%%%%%
\section{Calculation strategy}

To express a thermal state, we prepare a TPQ state~\cite{Sugiura_2012} on a quantum circuit.
The TPQ state, $|\psi \rangle^{\mathrm{TPQ}}$, is defined as a state that satisfies for low-degree polynomials of local operators, $^{\forall}\hat{O}$, and $\forall\epsilon>0$,
\begin{equation}
\mathrm{P}\left(\left|\frac{\langle \psi | \hat{O}| \psi \rangle^{\mathrm{TPQ}}}{\langle \psi | \psi \rangle^{\mathrm{TPQ}} }-\langle\hat{O}\rangle^{\mathrm{ens}}\right| \geq \epsilon\right) \leq \eta_{\epsilon}(N),
\end{equation}
at finite size system $N$.
Here $\langle\hat{O}\rangle^{\mathrm{ens}}$ presents the ensemble average and $\eta_{\epsilon}(N)$ is a function that vanishes at the thermodynamic limit ($N\rightarrow \infty$). 
We use a specific representation of the TPQ state termed as the canonical TPQ state~\cite{Sugiura_2013},
\begin{equation}
  |\beta, N\rangle \equiv e^{-\frac{\beta}{2} \hat{H}}\left|\psi_{R}\right\rangle,\label{eq:cTPQ}
\end{equation}
where $\beta$ denotes the inverse temperature, and the state $|\psi_{R}\rangle$ is random state.
Calculating the expectation value of the local operator $ \langle \hat{\mathcal{O}} \rangle^{\mathrm{TPQ}}_{\beta,N}  \coloneqq \langle \beta , N | \hat{\mathcal{O}} | \beta, N \rangle /\langle \beta,N |\beta,N \rangle$, then taking an average for initial random states, $\overline{\langle \hat{\mathcal{O}} \rangle_{\beta, N}^{\mathrm{TPQ}} }$, approaches the thermal ensemble average in the thermodynamic limit,
\begin{equation}
    \left(\overline{\langle\hat{\mathcal{O}}\rangle_{\beta, N}^{\mathrm{TPQ}}}-\langle\hat{\mathcal{O}}\rangle_{\beta, N}^{\mathrm{ens}}\right) \xrightarrow{N\rightarrow\infty} 0. \label{eq:prop-TPQ}
\end{equation}

To prepare the canonical TPQ state, Eq.~\eqref{eq:cTPQ}, on a quantum circuit, we first prepare the random state $|\psi_{R}\rangle$. Here, we use an algorithm proposed in Ref.~\cite{hunterjones2019unitary}, which incorporates a staggered arrangement of local unitary operations. 

The problem is how to perform the imaginary time evolution $e^{-\frac{\beta}{2} \hat{H}}$.
Here, we utilize the QITE algorithm~\cite{Motta_2019}, which is a quantum-classical hybrid algorithm designed to perform the imaginary time evolution on a quantum circuit.
In an infinitesimal time step ($\Delta \beta$) of the Trotter decomposition of a given Hamiltonian $\hat{H} = \sum_{l=1}^{L} \hat{h}[l]$, the imaginary time evolution, which is expressed by a non-unitary operator, can be approximated by a unitary operator, $ e^{-i \Delta \beta \hat{A}[l]}$.
The unitary operator can be expanded by the following set of unitary operators,
\begin{equation}
   e^{-i \Delta \beta \hat{A}[l]}= e^{-i \Delta \beta \sum_{I} a_{I}[l] \hat{\sigma}_{I}}.\label{eq:qite_unitary}
\end{equation}
Here,  $\hat{\sigma}_{I} \in\{I, X, Y, Z\}^{\otimes D}$ denotes the Pauli string, which acts on $D$-qubit domain.
 Since the index $I$ runs from $1$ to $4^D$, the coefficients $\{ a_{I}[l] \}$ form a real vector of size $4^D$.
 On the other hand, we suppose that a part of the Hamiltonian in each Trotter step, $\hat{h}[l]$, is a $k$-local operator. The domain size $D$ should be larger than $k$ for each Trotter step.
 In the case of a local Hamiltonian system, the algorithm works well even in $D\ll N $~\cite{Motta_2019}.
 In our Hamiltonian~\eqref{eq:hamiltonian_spin}, it includes all-to-all qubit interactions, so that we take $D=k(=N)$ in our calculation. 
 
Using the unitary operators, we obtain a state $|\Phi\rangle_{\text{unitary}} = e^{-i \Delta \beta \sum_{I} a_{I}[l] \hat{\sigma}_{I}}|\psi\rangle$. On the other hand, a desired state is $|\Phi\rangle_{\text{non-unitary}} = C e^{-\Delta \beta \hat{h}[l]}|\psi\rangle$, where $C$ is a normalization factor 
$C=\langle \psi |e^{-2\Delta\beta \hat{h}[l]}|  \psi\rangle$.
Here, we initialize the state $|\psi\rangle$ with a random state for our purpose of preparing the canonical TPQ state.
To find an optimized value of $a_I[l]$, we minimize the norm difference between these two states,
$\| \ |\Phi\rangle_{\text{unitary}}-|\Phi\rangle_{\text{non-unitary}}  \ \|,$  using classical computers.

In the process of fixing the value of $a_I[l]$, we calculate ($4^D \times 4^D$) matrix elements containing the expectation values of the Pauli strings. Ideally, the calculation of the expectation values is performed by quantum computers in the QITE algorithm. 
The bottleneck of this algorithm is the calculation of these numerous expectation values, in particular, for our non-local Hamiltonian case which requires a large value of $D$.
To address this computational obstacle, we developed a method to handle multiple Hamiltonian terms with different localities in a single Trotter step.
Furthermore, by utilizing the symmetry of Pauli strings, we reduce the computational cost for both quantum and classical calculations.
These improvements allow us to compute with large system sizes.

To obtain the thermodynamic quantity from the calculation data given by the QITE algorithm, we have to take double limits;
namely the $\Delta\beta \rightarrow 0$ limit  at finite $N$ to remove the Trotter errors and then $N\rightarrow \infty$.
After taking the first extrapolation, we can see that the result of QITE will be consistent with that of TPQ by classical calculation.
After the second extrapolation, we eventually obtain the value of thermodynamic quantity.

%%%%%%%%%%%%%%%%%%%%%%%%%%%%%%%%%%%%%%%%%%%%%%%%%%%%%%%%%%%%%%%%%%%%%%%%%%%
%%%%%%%%%%%%%%%%%%%%%%%%%%%%%%%%%%%%%%%%%%%%%%%%%%%%%%%%%%%%%%%%%%%%%%%%%%%
\section{Simulation results}
\subsection{Calculation methods and simulation parameters}
Now, we calculate the chiral condensate, Eq.~\eqref{eq:chiral_condensate}, as a function of temperature using three independent methods:
the exact diagonalization method, the ``TPQ method'', and the ``QITE method''.
In the second method, we classically prepare the TPQ state exactly as defined in Eq.~\eqref{eq:cTPQ} and calculate the expectation value of the chiral condensate.
Also in the third method, we generate the TPQ state, but here we perform the imaginary time evolution using the QITE algorithm.
This is a quantum-classical hybrid algorithm, but here we simulate the quantum part classically using a state vector simulator.

The lattice spacing and the coupling constant are set to $a=0.80$ and $g=1.00$, respectively. The lattice size $N$ is $4$--$12$.
The mass parameters in this work are massless and $m/g=0.15$. In the massive case, we also study the non-zero $\theta$ regime.
In both the TPQ and QITE methods, we utilize the same set of  $100$ initial random states and take the average over them.
Furthermore, in the QITE method, we take the Trotter step in the range of  $\Delta\beta =0.05$--$0.25$.

%%%%%%%%%%%%%%%%%%%%%%%%%%%%%%%%%%%%%%%%%%%%%%%%%%%%%%%%%%%%%%%%%%%%%%%%%%%
\subsection{Results of the massless case}
\begin{figure}[htbp]
  \begin{minipage}[b]{0.33\linewidth}
    \centering
    \includegraphics[keepaspectratio, scale=0.33]{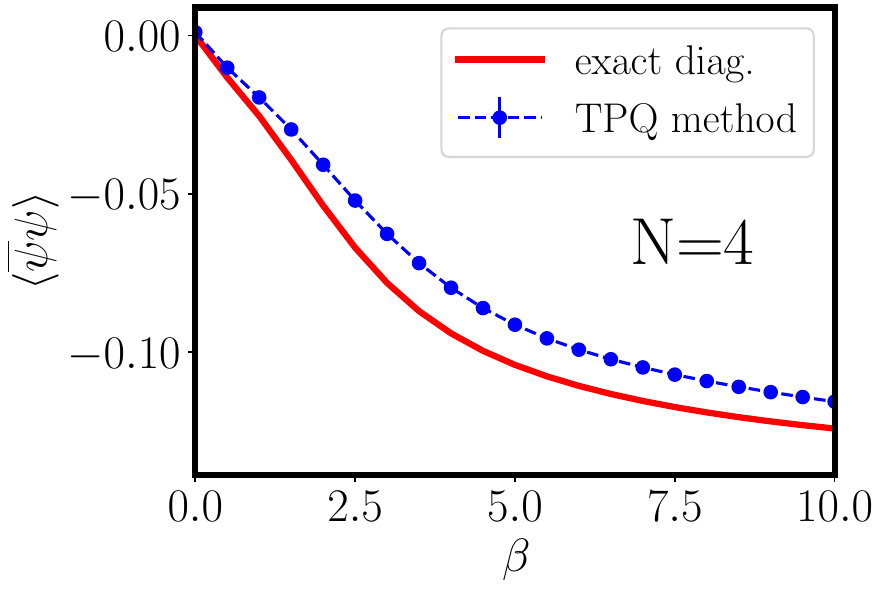}
  \end{minipage} 
  \begin{minipage}[b]{0.33\linewidth}
    \centering
    \includegraphics[keepaspectratio, scale=0.33]{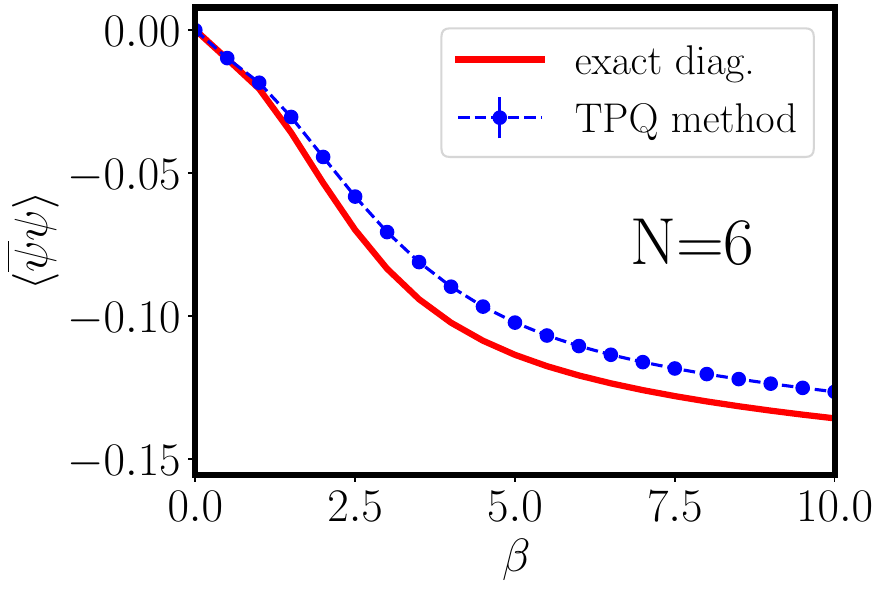}
  \end{minipage}
  \begin{minipage}[b]{0.33\linewidth}
    \centering
    \includegraphics[keepaspectratio, scale=0.33]{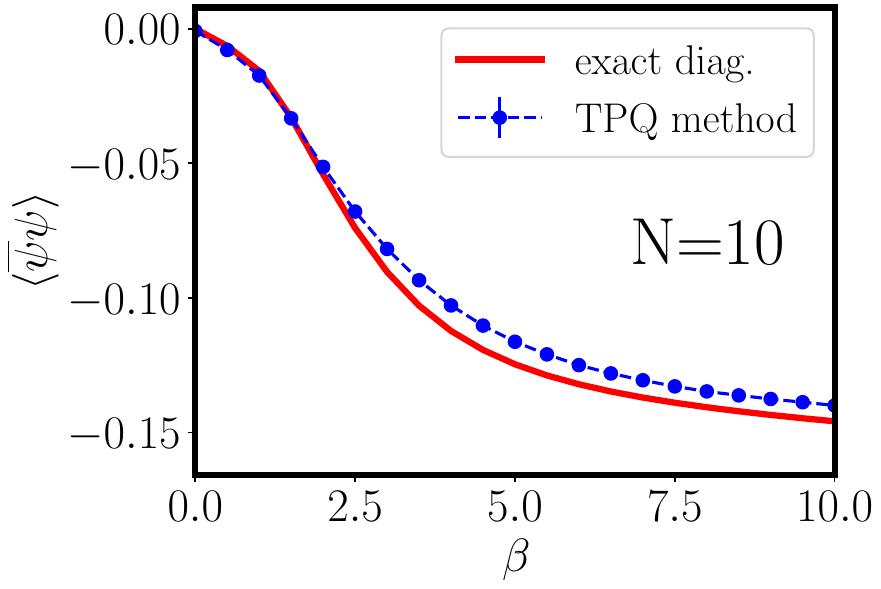}
  \end{minipage} 
  \caption{
  The comparison between the $\beta$-dependence of the chiral condensates calculated by the TPQ method (blue circle symbol) and the exact diagonalization (red curve) at finite $N$.
}\label{fig:massless_tpq_exact}
\end{figure}

First, we study the massless case as a feasibility test of our calculation strategy. In this case, the analytical result is known as shown in Eq.~\eqref{eq:exact_SW}.
Figure~\ref{fig:massless_tpq_exact} depicts the comparison between the results of the TPQ method (blue circle symbol) and the exact diagonalization results (red curve) at finite $N$.
%%%%%%%%%%%%%%%%%%%%%%%%%%%%%%%%%%%%%%%%%%%%%%%%%%%%%%%%%%%%%%%%%%%%%%%%%%%
%%%%%%%%%%%%%%%%%%%%%%%%%%%%%%%%%%%%%%%%%%%%%%%%%%%%%%%%%%%%%%%%%%%%%%%%%%%
The results of the TPQ method do not agree with the exact diagonalization results for each $N$, but the discrepancy becomes smaller in larger $N$.
We extrapolate these data at each $\beta$ at $N=4,6,8, 10$, and $12$ with the quadratic function and estimate the value at $N\rightarrow\infty$.

\begin{figure}[htbp]
\centering
\includegraphics[keepaspectratio, scale=0.5]{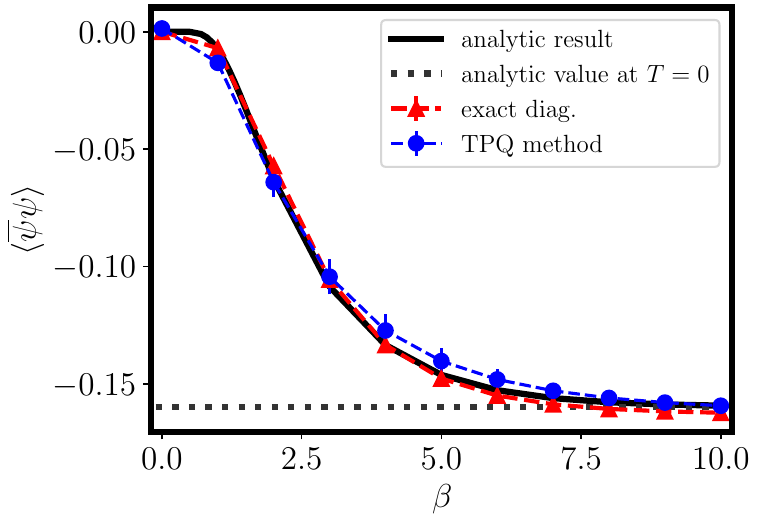}
\caption{
Comparison of the results in the thermodynamic limit obtained by the TPQ method (blue circle symbol), the exact diagonalization (red triangle symbol), and the analytical calculation, Eq.~\eqref{eq:exact_SW} (black solid curve).
The black dotted line shows the value at zero temperature~\cite{Gross_1996}.
}
\label{fig:massless_SW}
\end{figure}

Figure~\ref{fig:massless_SW} illustrates the $\beta$-dependence of the chiral condensate at the thermodynamic limit. Here, we plot the analytical result obtained by the Lagrangian formalism (black solid curve), the extrapolated value of the exact diagonalization results (red triangle symbol), and the extrapolated value of TPQ method (blue circle symbol).
The statistical error of the extrapolated values comes from the fitting error of the extrapolation.
These three results agree with each other within the statistical error as expected from Eq.~\eqref{eq:prop-TPQ}.

%%%%%%%%%%%%%%%%%%%%%%%%%%%%%%%%%%%%%%%%%%%%%%%%%%%%%%%%%%%%%%%%%%%%%%%%%%%
%%%%%%%%%%%%%%%%%%%%%%%%%%%%%%%%%%%%%%%%%%%%%%%%%%%%%%%%%%%%%%%%%%%%%%%%%%%

Furthermore, we can see that at $\beta=10$ the result converges to the value at zero temperature $\langle\bar{\psi} \psi\rangle \simeq-0.1599$~\cite{Gross_1996}. 
This indicates that $\beta = 10$ corresponds to a sufficiently low-temperature.

\begin{figure}[thbp]
  \begin{minipage}[b]{0.33\linewidth}
    \centering
    \includegraphics[keepaspectratio, scale=0.33]{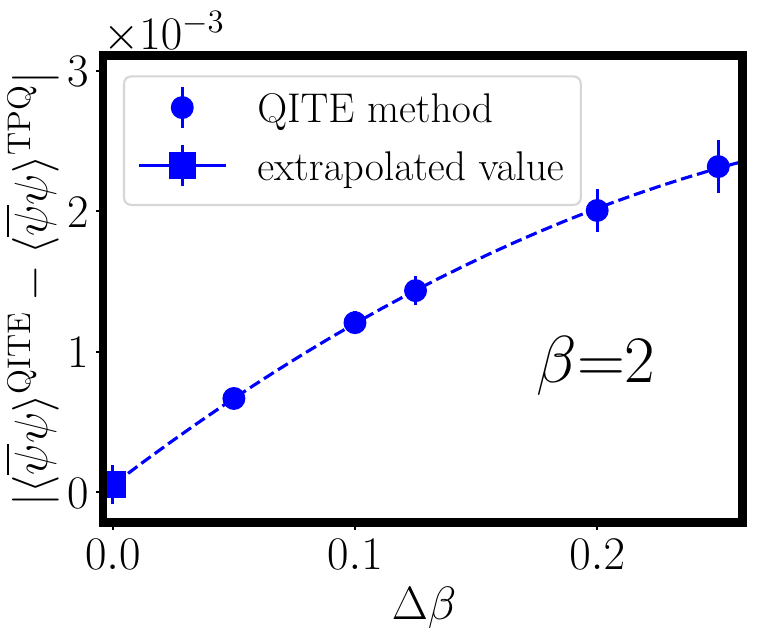}
  \end{minipage} 
  \begin{minipage}[b]{0.33\linewidth}
    \centering
    \includegraphics[keepaspectratio, scale=0.33]{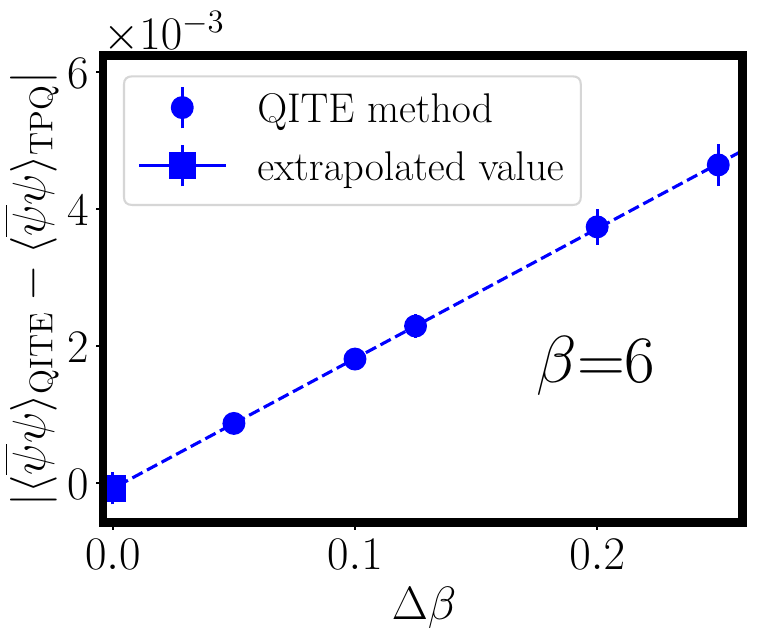}
  \end{minipage}
  \begin{minipage}[b]{0.33\linewidth}
    \centering
    \includegraphics[keepaspectratio, scale=0.33]{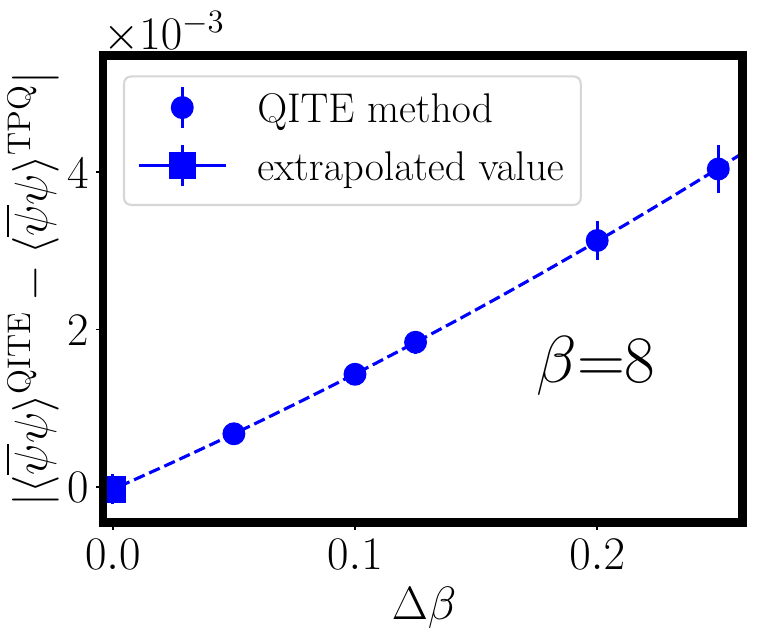}
  \end{minipage} 
  \caption{
  The $\Delta\beta$-dependence of the Trotter error in the QITE method.
  The blue circle symbol shows the absolute difference of the chiral condensate calculated by the TPQ and QITE methods and the blue square symbol shows the extrapolated value of the absolute difference in the $\Delta\beta\rightarrow 0$ limit.
}
\label{fig:massless_digi_error}
\end{figure}

Next, we examine the QITE method.
Here, we fix the lattice size as $N=6$. We take $\Delta\beta= 0.05, 0.10, 0.125, 0.20$, and $0.25$ and investigate the effects of the Trotter error.
In Figure~\ref{fig:massless_digi_error}, we plot the absolute difference between results by the QITE and TPQ methods as a function of the Trotter step size ($\Delta \beta$). It can be fitted well using the quadratic function of $\Delta \beta$ and the extrapolated value at $\Delta\beta\rightarrow 0$ is consistent with zero within the statistical error.
This result indicates that the Trotter error is under control and the QITE method reproduces the results of the TPQ method.
Combined with the consistency between the results of the TPQ method and the analytical results in the thermodynamic limit, our strategy based on TPQ and QITE looks promising to study the local observables at finite-temperature even for the Schwinger model Hamiltonian, which includes non-local terms.

%%%%%%%%%%%%%%%%%%%%%%%%%%%%%%%%%%%%%%%%%%%%%%%%%%%%%%%%%%%%%%%%%%%%%%%%%%%
\subsection{Results of the massive case}

\begin{figure}[htbp]
\centering
\includegraphics[keepaspectratio, scale=0.45]{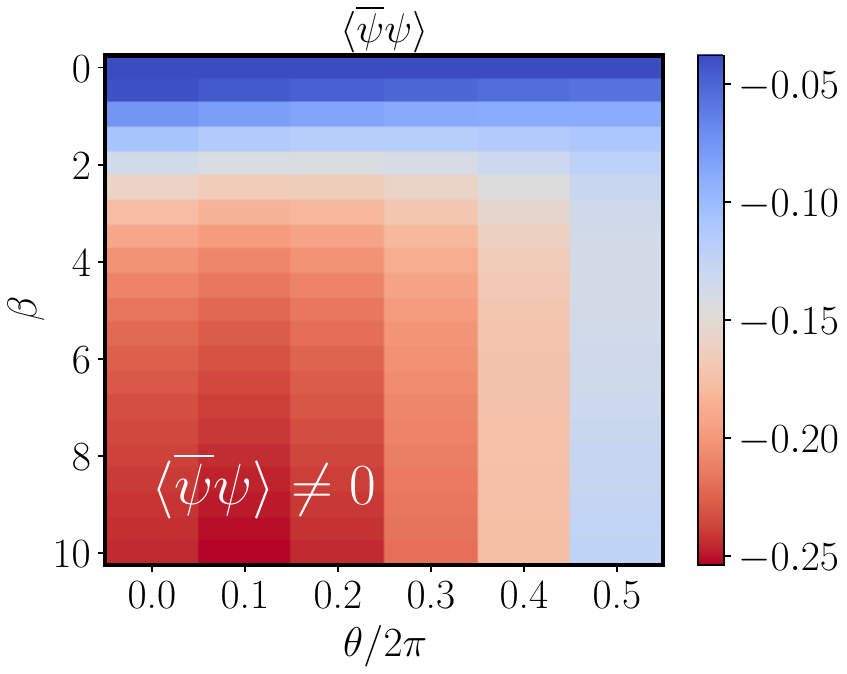}
\caption{
The density plot of the chiral condensate $\langle\bar{\psi}\psi\rangle$ with $N=6$, calculated by the QITE method. 
}
\label{fig:massive_density}
\end{figure}

We now consider the massive and finite $\theta$ regimes. 
Figure~\ref{fig:massive_density} shows the value of the chiral condensate computed by the QITE method in the $\beta$ -- $\theta$ plane.
Here, we take $m/g=0.15$ and $0 \leq \theta \leq \pi$ with the $ \theta/2\pi =0.1$ interval. 
%%%%%%%%%%%%%%%%%%%%%%%%%%%%%%%%%%%%%%%%%%%%%%%%%%%%%%%%%%%%%%%%%%%%%%%%%%%
%%%%%%%%%%%%%%%%%%%%%%%%%%%%%%%%%%%%%%%%%%%%%%%%%%%%%%%%%%%%%%%%%%%%%%%%%%%
We can see that the absolute value of the chiral condensate is large only in small $\theta$ and large $\beta$ regimes.
Outside this region, it gradually approaches zero, indicating that the (discrete) chiral symmetry is being restored.
The next task is the double extrapolation of $\Delta \beta \rightarrow 0 $ and $N \rightarrow \infty$, which is an ongoing work.

%%%%%%%%%%%%%%%%%%%%%%%%%%%%%%%%%%%%%%%%%%%%%%%%%%%%%%%%%%%%%%%%%%%%%%%%%%%
\section{Summary and discussion}
In this study, we simulated the Schwinger model at finite temperatures using the TPQ state approach and the QITE algorithm. 
As a feasibility test, we first considered the massless Schwinger model and showed that the (classical) TPQ method reproduces the result given by the exact diagonalization in the thermodynamic limit.
Furthermore, these results are consistent with the analytical result in the continuum limit based on the Lagrangian formalism.
We next confirmed that the results of the QITE method converge to those of the (classical) TPQ method in the Trotter error $\Delta \beta \rightarrow 0$ limit.
Thus, we found that the Trotter error for the discrete imaginary time evolution is under control in our calculation setup. 
We conclude that our method based on TPQ and QITE works well even for the Schwinger model Hamiltonian, which includes non-local terms.

We also simulated the massive Schwinger model with non-zero $\theta$-term. 
We explored the $\theta$ and $\beta$-dependence of the chiral condensate for a finite system.
The chiral condensate exhibits non-zero values in the region of small $\theta$ and high $\beta$, and gradually approaches zero outside this region. 
Taking both $\Delta \beta \rightarrow 0$ and $N\rightarrow \infty$  limits in the massive and non-zero $\theta$ case will be reported in the near future.

%%%%%%%%%%%%%%%%%%%%%%
\acknowledgments
We would like to thank K.~Nitadori for the useful discussions.
The numerical simulations are supported by the Yukawa-21 in YITP, Kyoto University and HOKUSAI supercomputer at RIKEN (Project ID No. Q22577).
The work of J.~W.~P is supported in part by the JSPS Grant-in-Aid for Research Fellow Number\ 22J14732 and the JST SPRING, Grant Number JPMJSP2108.
The work of E.~I. and S.~Y. is supported by 
JST Grant Number JPMJPF2221  % SQAI
and Program for Promoting Researches on the Supercomputer ``Fugaku'', Joint Institute for Computational Fundamental Science (JICFuS), Grant Number JPMXP1020230411.%Fugaku
The work of E.~I. is supported by JSPS KAKENHI with Grant Number 23H05439, %Kiban-S
JST PRESTO Grant Number JPMJPR2113, %Sakigake
and
JSPS Grant-in-Aid for Transformative Research Areas (A) JP21H05190. %ExU
The work of S.~Y. is supported by 
JSPS Grant-in-Aid for Scientific Research (A) JP21H03455, 
and the COE research grant in computational science from Hyogo Prefecture and Kobe City through Foundation for Computational Science.

% The numerical simulations are supported by the Yukawa-21 in YITP, Kyoto University and HOKUSAI in RIKEN.
% The work of J.~W.~P is supported in part by the JSPS Grant-in-Aid for Research Fellow Number\ 22J14732 and the JST SPRING, Grant Number JPMJSP2108.
% The work of E.~I. is supported by JSPS KAKENHI with Grant Number 23H05439,%Kiban-S
% JST PRESTO Grant Number JPMJPR2113,%Sakigake
% JSPS Grant-in-Aid for Transformative Research Areas (A) JP21H05190, %ExU
% JST Grant Number JPMJPF2221  % SQAI
% and also supported by Program for Promoting Researches on the Supercomputer ``Fugaku'' (Simulation for basic science: from fundamental laws of particles to creation of nuclei) and (Simulation for basic science: approaching the new quantum era), and Joint Institute for Computational Fundamental Science (JICFuS), Grant Number JPMXP1020230411.%Fugaku

\bibliographystyle{utphys}
\bibliography{LATTICE23,./Nf2_Schwinger.bib,./QFT.bib}

\providecommand{\href}[2]{#2}\begingroup\raggedright\begin{thebibliography}{10}\setlength{\itemsep}{-2pt}

\bibitem{Sugiura_2012}
S.~Sugiura and A.~Shimizu \href{http://dx.doi.org/10.1103/physrevlett.108.240401}{{\em Physical Review Letters} {\bfseries 108} no.~24, (Jun, 2012) }.

\bibitem{Sugiura_2013}
S.~Sugiura and A.~Shimizu \href{http://dx.doi.org/10.1103/physrevlett.111.010401}{{\em Physical Review Letters} {\bfseries 111} no.~1, (Jul, 2013) }.

\bibitem{davoudi2022quantum}
Z.~Davoudi, N.~Mueller, and C.~Powers \href{http://arxiv.org/abs/2208.13112}{{\ttfamily arXiv:2208.13112 [hep-lat]}}.

\bibitem{McArdle_2019}
S.~McArdle, T.~Jones, S.~Endo, Y.~Li, S.~C. Benjamin, and X.~Yuan \href{http://dx.doi.org/10.1038/s41534-019-0187-2}{{\em npj Quantum Information} {\bfseries 5} no.~1, (Sep, 2019) }.

\bibitem{Liu_2021}
T.~Liu, J.-G. Liu, and H.~Fan \href{http://dx.doi.org/10.1007/s11128-021-03145-6}{{\em Quantum Information Processing} {\bfseries 20} no.~6, (Jun, 2021) }.

\bibitem{seki2021}
K.~Seki and S.~Yunoki \href{http://dx.doi.org/10.1103/PRXQuantum.2.010333}{{\em PRX Quantum} {\bfseries 2} (Feb, 2021) 010333}.

\bibitem{Motta_2019}
M.~Motta, C.~Sun, A.~T.~K. Tan, M.~J. O'Rourke, E.~Ye, A.~J. Minnich, F.~G. S.~L. Brand{\~{a}}o, and G.~K.-L. Chan \href{http://dx.doi.org/10.1038/s41567-019-0704-4}{{\em Nature Physics} {\bfseries 16} no.~2, (Nov, 2019) 205--210}.

\bibitem{Honda:2021aum}
M.~Honda, E.~Itou, Y.~Kikuchi, L.~Nagano, and T.~Okuda \href{http://dx.doi.org/10.1103/PhysRevD.105.014504}{{\em Phys. Rev. D} {\bfseries 105} no.~1, (2022) 014504}, \href{http://arxiv.org/abs/2105.03276}{{\ttfamily arXiv:2105.03276 [hep-lat]}}.

\bibitem{sachs2010finite}
I.~Sachs and A.~Wipf \href{http://arxiv.org/abs/1005.1822}{{\ttfamily arXiv:1005.1822 [hep-th]}}.

\bibitem{Gross_1996}
D.~J. Gross, I.~R. Klebanov, A.~V. Matytsin, and A.~V. Smilga \href{http://dx.doi.org/10.1016/0550-3213(95)00655-9}{{\em Nuclear Physics B} {\bfseries 461} no.~1-2, (Feb, 1996) 109--130}.

\bibitem{hunterjones2019unitary}
N.~Hunter-Jones \href{http://arxiv.org/abs/1905.12053}{{\ttfamily arXiv:1905.12053 [quant-ph]}}.

\end{thebibliography}\endgroup

\end{document}